\documentclass{PoS}
\newcommand{\HG}{\hat{\Gamma}}
\newcommand{\be}{\begin{equation}}
\newcommand{\ee}{\end{equation}}
\newcommand{\ba}{\begin{array}}
\newcommand{\ea}{\end{array}}
\newcommand{\baa}{\begin{array}}
\newcommand{\eaa}{\end{array}}
\newcommand{\bea}{\begin{eqnarray}}
\newcommand{\eea}{\end{eqnarray}}

\newcommand{\pc} {{\vec p\, }^c}

\title{Volume dependence in 2+1 Yang-Mills theory}

\ShortTitle{Volume dependence in 2+1 Yang-Mills theory}

\author{\speaker{Margarita Garc\'{\i}a P\'erez}%
         \\
        Instituto de F\'{\i}sica Te\'orica UAM/CSIC, Universidad Aut\'onoma de Madrid, 
E-28048-Madrid, Spain\\
        E-mail: \email{margarita.garcia@uam.es}}

\author{Antonio Gonz\'alez-Arroyo\\
Instituto de F\'{\i}sica Te\'orica UAM/CSIC and  Departamento de F\'{\i}sica Te\'orica, C-15,
Universidad Aut\'onoma de Madrid, E-28049-Madrid, Spain \\
        E-mail: \email{antonio.gonzalez-arroyo@uam.es}}

\author{Masanori Okawa\\
Graduate School of Science, Hiroshima University, 
Higashi-Hiroshima, Hiroshima 739-8526, Japan\\
        E-mail: \email{okawa@sci.hiroshima-u.ac.jp}}

\abstract{We present the results of an analysis of a 2+1 dimensional pure SU($N$) 
Yang-Mills theory formulated on a 2-dimensional spatial torus with non-trivial 
magnetic flux. 
We focus on investigating the dependence of the electric-flux spectrum, 
extracted from Polyakov loop correlators, with the spatial size $l$, 
the number of colours $N$, and the magnetic flux $m$. The size of the torus 
acts a parameter that allows to control the onset of non-perturbative effects. 
In the small volume regime, 
where perturbation theory holds, we derive the one-loop self-energy correction 
to the single-gluon spectrum, for arbitrary $N$ and $m$. We discuss the 
transition from small to large volumes that has been investigated by means of 
Monte-Carlo simulations.
We argue that the energy of electric flux $\vec e$, for the lowest gluon momentum,
depends solely on $\vec e /N$ and the dimensionless variable $x=\lambda N l$, 
with $\lambda=g^2 N$ the 't Hooft coupling. The variable $x$ can be interpreted 
as the dimensionless 't Hooft coupling for an effective box size given by $Nl$. 
This implies a version of reduction that allows to trade $l$ by  $N$ 
without modifying the electric-flux energy.}

\FullConference{The 30th International Symposium on Lattice Field Theory\\
                 June 24 - 29,  2012\\
                 Cairns, Australia}

\begin{document}

\section{Introduction}
\label{s:intro}
In this paper we will explore the electric-flux spectrum
of a 2+1-dimensional SU(N) Yang-Mills theory defined on a 2-dimensional spatial 
torus endowed with a chromo-magnetic flux.  The aim is to disentangle the 
dependence on the 
number of colours $N$ and the size of the torus $l$, following the idea of reduction
introduced by Eguchi and Kawai~\cite{EK,GonzalezArroyo:2010ss}.  
The size of the torus will
act as a control parameter that permits to explore the onset of non-perturbative 
dynamics. This is so, because for small $l$ the effective coupling 
constant becomes small and perturbation theory holds. As 
$l$ grows, the finite size effects, including the magnetic flux, should become 
irrelevant. The way in which this takes place, and in particular the interplay 
with the large $N$ limit, might shed some light into the processes relevant 
for non-perturbative dynamics. 
We will focus here for simplicity on the 2+1 dimensional 
case that shares many of the non-perturbative properties of the 4-dimensional theory.
There is an extensive literature on the subject of Yang-Mills 3-d fields and large $N$ 
dynamics, for recent lattice reviews we refer the reader to \cite{Narayanan:2007fb}.

\section{Set-up and Perturbative Analysis}
\label{s:pert}

We will be considering SU($N$) Yang-Mills theories 
defined on a spatial torus of size $l\times l$. In the basis of constant transition
matrices $\Gamma_i$, the gauge field connection has to 
satisfy the periodicity condition:
$A_i(x+ l \hat \jmath)=\Gamma_j A_i(x) \Gamma^\dagger_j$ ,
with the $\Gamma_i$ fulfilling:
\begin{equation}
\label{MCOM}
\Gamma_1 \Gamma_2 = \exp \Big \{ i \, {2 \pi m \over N}\Big \}\, \, 
\Gamma_2 \Gamma_1\quad ,
\end{equation}
where $m$ is the magnetic flux. In the case that $m$ and $N$ are
co-prime, this equation defines the matrices $\Gamma_i$ uniquely
modulo global gauge transformations.
For simplicity we will assume that $N$ is odd and co-prime with $m$ in the following.
The periodicity constraint on the gauge fields can be solved by 
introducing a basis of $N\times N$ matrices $\HG(\pc)$ satisfying:
\begin{equation}
\Gamma_i\HG(\pc)\Gamma_i^\dagger = e^{i l {p}_i^{\, c}} 
\HG(\pc) \quad ,
\end{equation}
where $\pc =\frac{2 \pi \vec n}{l N}$,
with $n_i$  integers  defined modulo N. Thus, there are
$N^2$ such matrices.  An explicit solution to the equation is:
\be
\label{defHG}
\HG(\pc) = e^{i \alpha(\pc)} \, \, \Gamma_1^{\, -\, \bar{k} n_2} \, \Gamma_2^{\, \bar{k}\,  n_1} \quad ,
\ee
where $\bar{k}$ is an integer satisfying $ m \bar{k} = 1 \bmod N$,
and $\alpha(\pc)$ an arbitrary phase factor.
In this basis we can Fourier decompose the gauge connection as:
\be
\label{decomposition}
A_i(x)=  {\cal N}\, \sum'_{\vec{P}} \, e^{i \vec{P}\vec{x}}\, \hat{A}_i(\vec{P})\, 
 \HG(\vec{P}) \quad .
\ee
The total momentum $\vec{P}$ is quantized in units of $2 \pi /(Nl)$ and can
be decomposed as: $\vec{P}= \vec{p}+ \pc$, with $\vec p=2 \pi \vec n/l$, 
and $n_i \in Z\!\!\!Z$.
The prime means that $\pc= \vec 0$ is excluded from the sum, since
$A_i(x)$ has to be traceless.
This can be interpreted by saying that the total momentum $\vec{P}$ is
composed of two pieces: a colour-momentum part $\pc$ and 
a spatial-momentum part $\vec{p}$.
If we neglect the prime in the sum, the range of values of
$\vec{P}$ is just that of a theory defined on a box of size $Nl\times Nl$. 
The formalism looks hence as if the theory had no colour, but was
defined on a bigger spatial box with effective size $Nl$.

We will now analyze in perturbation theory the energy spectrum of one-gluon states.
The first remark is that, since zero momentum is excluded, the theory has
a mass-gap even for small $N l$. At tree level in perturbation theory, the mass
of a state with colour momentum $\pc$ is   
$E_0=|\pc|=2 \pi |\vec n|/(Nl)$. Such states can be characterized in a gauge-invariant way in
terms of the electric flux $\vec e$, which is defined modulo $N$ and related 
to $\vec n$ through: $e_i = \epsilon_{j i} n_j \bar k$. The gauge invariant operator 
projecting onto a state of electric flux $\vec e$ is a Polyakov loop winding 
$e_i$ times around each of the cycles of the 2-torus. The non-perturbative spectrum
of such states can hence be obtained from Polyakov loop correlators.

\begin{figure}
\begin{center}
\includegraphics[scale=0.26,angle=-90]{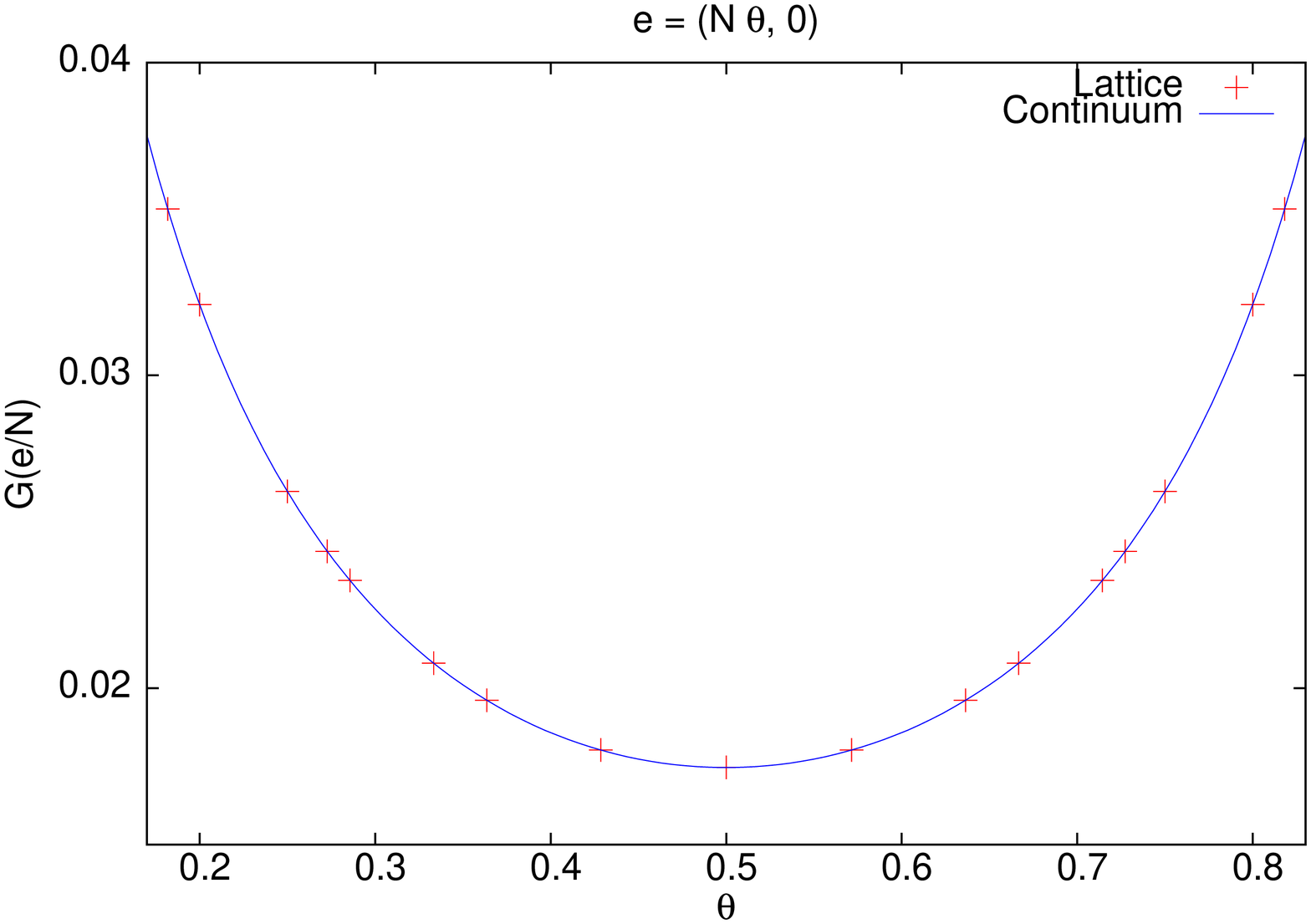}
\includegraphics[scale=0.26,angle=-90]{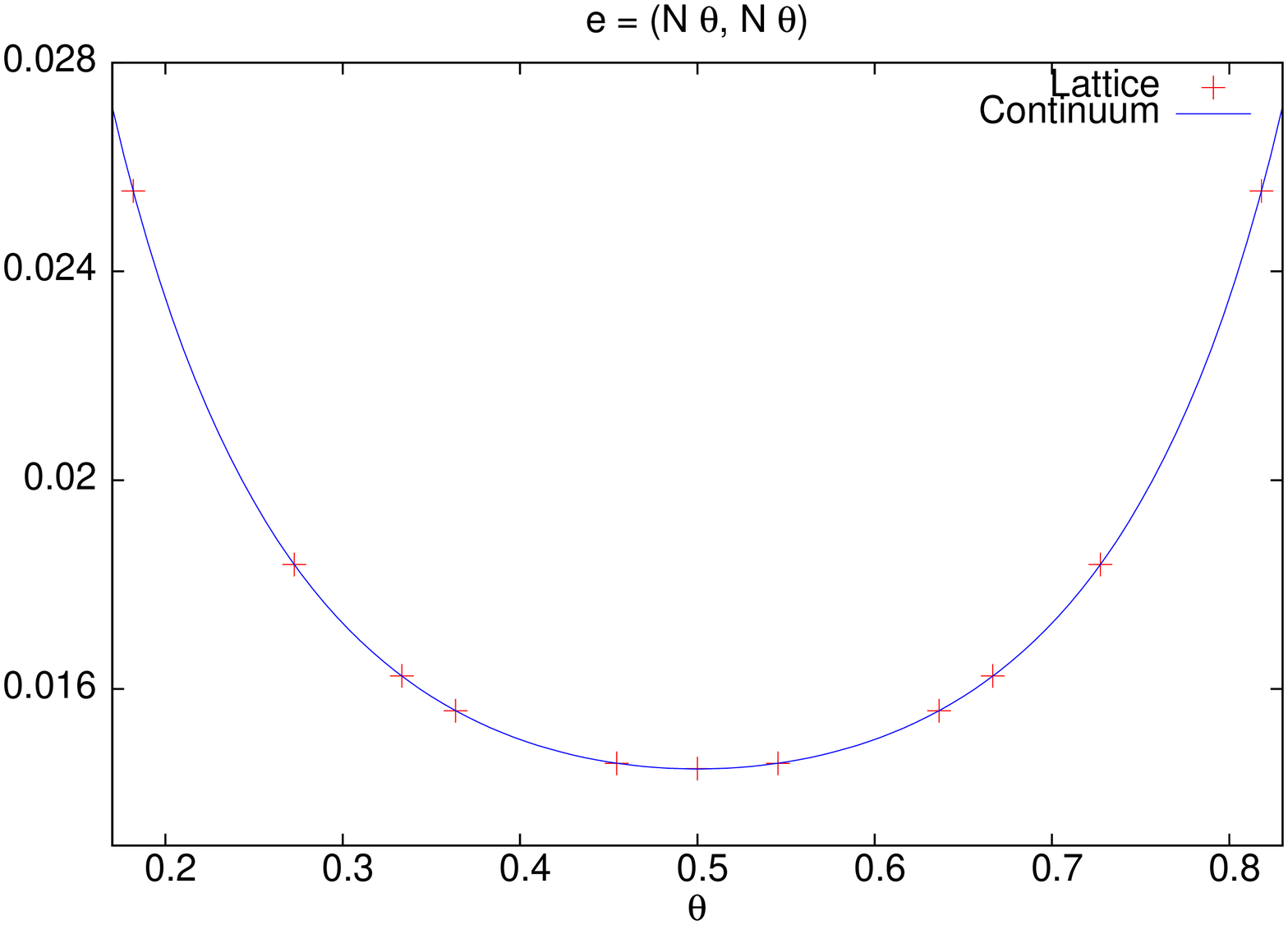}
\caption{\label{f:G}
We show, for electric flux $\vec e =(N\theta, 0)$ (left), 
and $\vec e =(N\theta,N\theta)$ (right), the function $G(\theta)$ that  
gives the one-loop correction to the energy of a one-gluon state, through Eq.~(2.5). 
The blue line corresponds 
to Eq.~(2.6), while the red points are derived using a lattice regularization 
in the calculation of the self-energy.}
\end{center}
\end{figure}

 In perturbation theory the energy of electric flux can be determined by computing
the gluon self-energy on the twisted box (for examples in SU(2) see Refs.~\cite{GA-KA,daniel}). Details of the calculation at one-loop order will be presented 
elsewhere~\cite{mam}. Here we simply provide the final expression that will be used
to analyze the numerical results in section \ref{s:num}.   
Putting everything together, we obtain for the square of the energy of electric flux
$\vec e$ with colour momentum $\vec n$:
\be
\label{eq:Esq}
{E^2(\vec e) \over \lambda^2}=  \Big ( { 2 \pi |\vec n| \over  \lambda N l }\Big )^2 \,
-  \,{ 4 \pi \over  \lambda N l} \,  G \Big ({\vec e \over N}\Big ) \quad ,
\ee
where $\lambda = g^2 N$ is the 't Hooft coupling, which is dimensionful in 2+1 
dimensions. The first order correction to the tree level expression is given in terms of: 
\be
\label{eq:G}
G \Big ({\vec e \over N}\Big) = -{ 1 \over 16 \pi^2 } \int_0^{\infty} {dx \over \sqrt{ x}}
\, \Big(\theta_3^2(0,ix) - \prod_{i=1}^2\, \theta_3({e_i\over N},i x)- {1 \over x}\Big)
\ee
with the Jacobi Theta function given by:
\be
\theta_3(z,ix) = \sum_{n \in {\bf Z}} \exp\{-x \pi n^2 + 2 \pi i n z\}\quad .
\ee
The dependence of $G$ on $\vec e /N$ is exhibited in 
Fig.~\ref{f:G}, for electric fluxes of the form $\vec e = (N \theta,0)$ and
$\vec e = (N \theta,N \theta)$.

One important remark at this stage is that, once $\vec n$ and  
$\vec e/N$ have been fixed, the energy depends 
exclusively on the variable $x= \lambda N l$, 
which can be interpreted as the dimensionless 't Hooft coupling corresponding to an 
effective box of size $Nl$. Based on our numerical results, we will argue that this is
the case even in the non-perturbative regime. This will imply thus a kind of reduction 
in which $N$ and $l$ become interchangeable.
A second remark in order is the non-analyticity of $G(\theta)$ at 
$\theta= 0 \, ({\rm mod} \, 1)$. This could give rise to a tachyonic instability in 
the energy of electric flux for $N\rightarrow \infty$, which  
has been previously discussed in the context of non-commutative geometry in 
Ref.~\cite{Guralnik:2002ru}. There it was presented as an evidence of 
spontaneous $Z_N$ symmetry breaking.
We will come back to this issue when discussing the numerical results in 
section~\ref{s:disc}.

Let us finally mention that in order to arrive at the one-loop perturbative 
expression we have used three different approaches: to 
compute the Euclidean gluon self-energy in the continuum and on the lattice, 
and to use the Hamiltonian formulation. The lattice calculation has made use of 
the results in Refs. \cite{Snippe:1996bk},\cite{Luscher:1984xn}.
In Fig.~\ref{f:G}, the lattice results for $G(\theta)$, extrapolated to the continuum 
limit, are compared with the formula given in Eq. (\ref{eq:G}).  
There is perfect agreement between the two.

\section{Numerical Results}
\label{s:num}

In this section we will present the results of a lattice calculation of the
electric-flux spectrum  in a twisted box as a function of $N$, $l$ and the magnetic flux.
We start with a $L_0 \times L^2$ lattice with twisted boundary conditions ($l=La$, 
with $a$ the lattice spacing). 
It is possible to perform a change of variables of the link matrices that 
allows to work with periodic links at the price of introducing a twist 
dependent factor in the action: 
\be
S_W = N b \sum_{x}  \sum_{\mu \ne \nu} \Big ( N - z_{\mu \nu}^*(x)
\, {\rm Tr} P_{\mu \nu } (x) \Big ) \quad ,
\label{eq.wilson2}
\ee
where $z_{\mu \nu}(x)=1$  except for the corner plaquettes in each (1,2)
plane where it takes the value:
\be
z_{i j}(x) = \exp \Big \{i \, {2 \pi m\over N} \, \epsilon_{ij}\Big \} \quad ,
\ee
with $m$ the magnetic flux. We have performed Monte Carlo simulations with this 
action employing a heat-bath algorithm based on a proposal by Fabricius and 
Haan~\cite{Fabricius:1984wp}.
Four sets of $(N,L,L_0)$ values have been generated: (5,14,48), (5,22,72), (7,10,32), and (17,4,32),
with values of the magnetic flux:
$m=1,\, 2$, for SU(5),  $m=1,\, 2,\, 3$, for SU(7), 
and $m=1,\, 3,\, 5,\, 8$, for SU(17).

The electric-flux spectrum has been extracted from spatially smeared 
Polyakov loop correlators. The spatial Polyakov loops, winding $e$ times around
the $\hat \imath$th direction of the torus, read:
\be
P_i (t,\, \vec x,\,  e)= {\rm Tr} \Big\{ \prod_{s=0}^{L-1}  U_i \, (t,\,  \vec x+s \hat \imath)
\Big\}^{e}\quad .
\ee
The twist gives rise to non-trivial boundary conditions for the loops:
$P_i (t,\, \vec x+L \hat \jmath, \, e) = z_{ij}^{\, e} \,  \, P_i (t,\, \vec x, \, e)$,
that have to be considered when
extracting the spectrum from the Polyakov loop correlators. In practice, we project over
the different colour-momentum states by computing the averaged correlators:
\be
C_1(t,\,  e,\, n) = {1\over L^2} \sum_{y, \tilde y =1}^{L}  \exp \Big \{- i \, { 2 \pi n  \over L N} \, 
\tilde y \Big\} \, \langle \, P_1( 0,\,  y \hat \jmath,\, e)
\, P_1^\dagger(t,\,  (y + \tilde y )\hat \jmath,\, e) \, \rangle \quad ,
\ee
with $e=   \epsilon_{21}\, n\,\bar k $. An analogous expression holds for 
$C_2(t,\,  e,\, n)$. 
The energy is extracted from the exponential decay at large $t$ of the correlator.

\begin{figure}
\begin{center}
\includegraphics[scale=0.56,angle=-90]{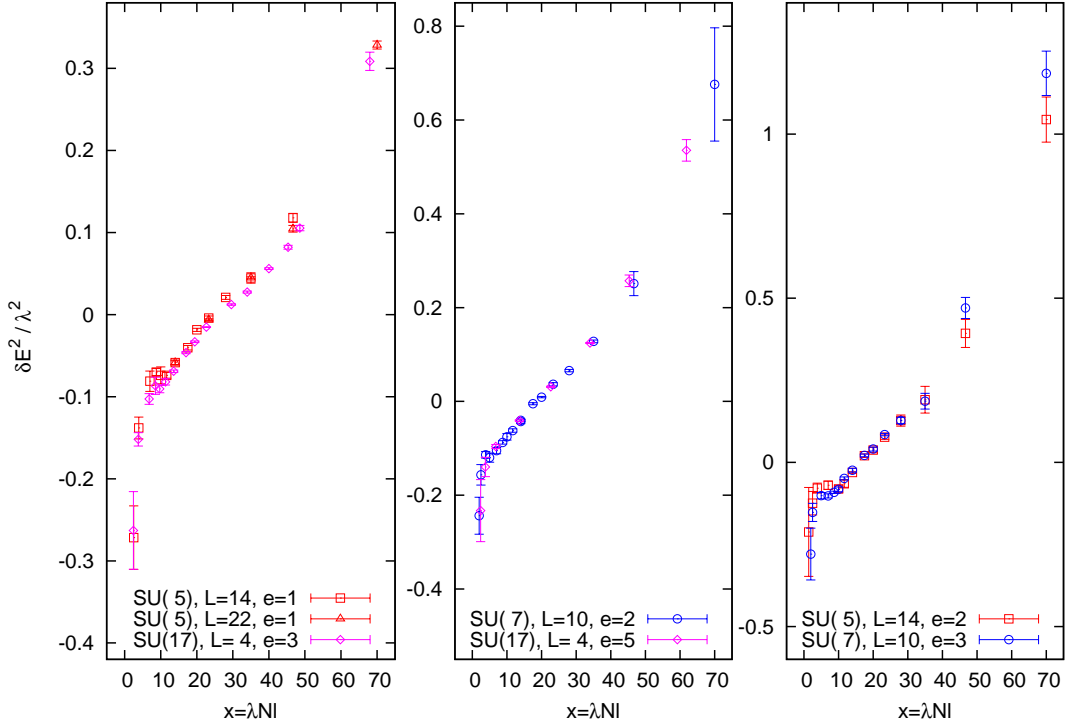}
\caption{\label{f:ener} For different values of $l$, $N$, and $e$,
we display the quantity
$\delta E^2(e) / \lambda^2 = (E^2 - E_0^2)/\lambda^2$ 
vs $x=\lambda N l$, where $
E$ is the energy of electrix flux $e$ with momentum $|\vec n| = 1$, and
$E_0= 2 \ {\rm asinh}(\sin(\pi /(NL)))$ is the lattice tree-level perturbative result.
}
\end{center}
\end{figure}


A detailed account of the results will be presented in \cite{mam}. Here we 
will focus on a few examples that illustrate the scaling of the 
electric-flux energies with $x=\lambda N l$. We will focus on the 
minimal momentum states with $|\vec n| = 1$ which have electric flux 
$|\vec e| = \bar k$.  According to our perturbative 
calculation, for given $e/N$ we expect energies to depend solely on $x$. 
Given that $N$ is prime, 
identical ratios of $ e / N$ are not possible, nevertheless our data 
falls in three sets with similar values for the ratio:   $(N,e)= \{(5,1), 
(17,3)\}; \, \{(7,2),(17,5)\}; \, \{(5,2), (7,3)\}$. The 
dependence of the energy squared on $x$ is presented in Fig.~\ref{f:ener}.
The scaling holds in all the range of $x$ analyzed which goes
far beyond the perturbative regime. A discussion of the results will be presented 
below.

\begin{figure}
\begin{center}
\includegraphics[scale=0.56,angle=-90]{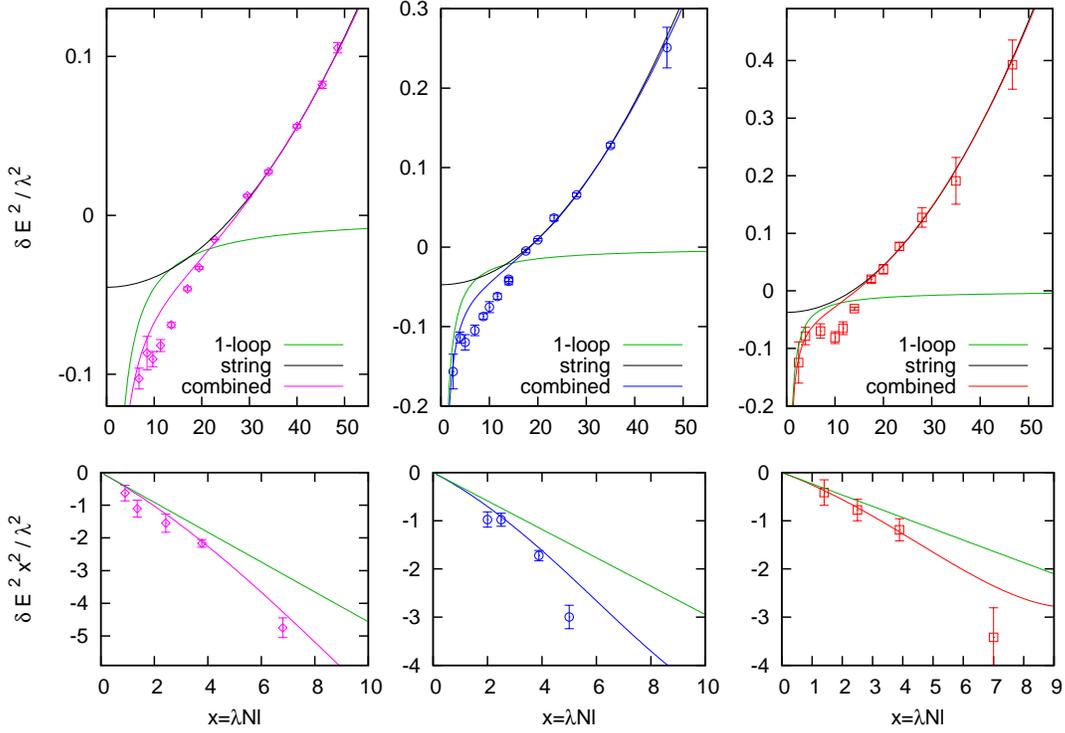}
\caption{\label{f:fits} In the upper panels we display
several fits of $\delta E^2(e) / \lambda^2$ to Eq.~(4.4). The symbol 
codes are the same as in Fig. 2.
Different lines correspond to:
(combined) $\alpha= G(e/N)$, $\beta$ and $\gamma$ free;
(1-loop) $\alpha= G(e/N)$, $\beta=\gamma=0$;
(string) $\alpha=0$, $\beta$ and $\gamma$ free.
The lower panels display $\delta E^2(e) x^2/ \lambda^2$.
}
\end{center}
\end{figure}

\section{Discussion}
\label{s:disc}

Our numerical results indicate that the energy of electric flux, in units of the
't Hooft coupling, depends solely on the variable $x=\lambda Nl$ and the value of 
$e/N$, at least for the minimal momentum $|\vec n|=1$.
This does indeed hold for the perturbative expression.
Let us look now at the expectation in the
large $x$ regime, where strings of electric flux are expected to be formed and the 
energy should grow linearly with the box size $l$. A large amount of lattice results 
hint in the direction that flux tubes can be described in terms of an effective string
picture based on the Nambu-Goto action~\cite{Narayanan:2007fb}. In the presence of 
a constant background magnetic field $B$, the Nambu-Goto prediction for a closed-string 
winding $\vec e$ times around the torus is given by:
\be
{E^2(\vec e) \over \lambda^2} =  \Big ( {\sigma |\vec e|  \over \lambda^2 N}\Big )^2 (\lambda N l)^2
 - {\pi \sigma \over 3 \lambda^2} + \sum_i \Big ( { \epsilon_{ij} e_j \, B \over 
\lambda l }\Big)^2 \quad ,
\ee
where $\sigma$ is the fundamental string tension. The last term on the right hand 
side of this formula can be easily mapped onto the tree-level 
perturbative expression $E_0^2$ by taking into account the relation between 
$\vec n$ and $\vec e$, and identifying the magnetic field with $B= 2 \pi m /N$. 
If the Nambu-Goto expression would hold, two different regimes
would take place in terms of the scaling variable $x$:
\begin{itemize}
\item Low $x$, where:
\be
{E^2(\vec e) \over \lambda^2} - {E_0^2(\vec e) \over \lambda^2}=  -  \,{ 4 \pi \over x} \,  
G \Big({\vec e \over N}\Big) \quad .
\ee
\item Large $x$, where:
\be
{E^2(\vec e)  \over \lambda^2} - {E_0^2(\vec e) \over \lambda^2}=  
\Big ( {\sigma |\vec e|  \over \lambda^2 N}\Big )^2 x^2 - {\pi \sigma \over 3 \lambda^2}\quad .
\ee
\end{itemize}
This would be fully consistent with our hypothesis on the $x$ scaling, once $\vec e/N$ 
is fixed. 
A formula that captures the asymptotic behaviour in both the low and large $x$ regimes
is:
\be
\label{eq:fits}
{\delta E^2(e) \over \lambda^2} = {E^2(e)  \over \lambda^2} - {E_0^2(e) \over \lambda^2}= -{4 \pi \alpha \over x} 
- {\pi \beta \over 3} +\gamma^2 \, x^2
\quad .
\ee
Fitting this expression to our data gives the results  
displayed in Fig.~\ref{f:fits}. The fits are restricted to the large $x$ region 
with $x > 20$ ($x > 30$, for $N=17$, $e=3$). 
Two different fits are presented: (a) fixing $\alpha=G(e/N)$, denoted by 
`combined', and (b) fixing $\alpha=0$, denoted by `string'.
In all cases, the `combined' fit has a slightly better $\chi^2$ per degree of 
freedom than the `string' fit. In addition, we display the 1-loop expression.
Although the combined formula fails to reproduce the results in the intermediate 
$x$ region it describes qualitatively quite well the data. 
One remarkable observation is 
that for small $x$ it improves the one-loop result in the correct direction. 
This is clearly observed in the low panels of Fig.~\ref{f:fits} where $\delta E^2 x^2
/\lambda^2$ is plotted to enhance this effect. We are at present exploring alternative 
parametrizations to improve the fit in the intermediate $x$ regime. 

One final remark refers to the possible existence of tachyonic instabilities.
One could use Eq.~(\ref{eq:fits}) to see in which instances the energy remains 
non-tachyonic for all values of $x$. 
A more thorough analysis will be presented in \cite{mam}. Let us just mention here that
our results indicate that it suffices to keep both $m$ and $\bar k$ of order 
$N$ to avoid the tachyonic behaviour. This coincides with the criteria introduced in 
\cite{GonzalezArroyo:2010ss} for twisted Eguchi-Kawai reduction to hold at large $N$.

\acknowledgments 
M.G.P. and A.G-A acknowledge financial support from the MCINN grants FPA2009-08785 and 
FPA2009-09017, the Comunidad Aut\'onoma
de Madrid under the program  HEPHACOS S2009/ESP-1473, and the European Union under
Grant Agreement number PITN-GA-2009-238353 (ITN STRONGnet).
They participate in the Consolider-Ingenio 2010 CPAN 
(CSD2007-00042). M. O. is supported by the Japanese MEXT grant No 23540310. 
We acknowledge the use of the IFT clusters for our numerical
results.

\end{document}